# Real-time digital signal recovery for a multi-pole low-pass transfer function system

Jhinhwan Lee[1,a)]

[1]Department of Physics, Korea Advanced Institute of Science and Technology, Daejeon 34141, Korea

In order to solve the problems of waveform distortion and signal delay by many physical and electrical systems with multi-pole linear low-pass transfer characteristics, a simple digital-signal-processing (DSP)-based method of real-time recovery of the original source waveform from the distorted output waveform is proposed. A mathematical analysis on the convolution kernel representation of the single-pole low-pass transfer function shows that the original source waveform can be accurately recovered in real time using a particular moving average algorithm applied on the input stream of the distorted waveform, which can also significantly reduce the overall delay time constant. This method is generalized for multi-pole low-pass systems and has noise characteristics of the inverse of the low-pass filter characteristics. This method can be applied to most sensors and amplifiers operating close to their frequency response limits to improve the overall performances of data acquisition systems and digital feedback control systems.

## Introduction

All first-order-response sensors and amplifiers suffer from delayed and distorted responses with single-pole low-pass characteristics when operated close to their frequency response limits. Notable examples include thermometers with finite heat capacity and thermal resistance, electro-mechanical sensors such as MEMS-based pressure and acceleration sensors with finite inertia of the moving parts, and electrometer-grade current amplifiers with a large feedback resistor with a finite shunt capacitance. In these systems, in order to achieve minimally distorted waveform and negligible phase delay often required for feedback control applications, the change in the input signal needs to be slower than the time constant of the first-order system by one or two orders of magnitude, which is not desirable in many speed-demanding applications. Here I propose a novel real-time numerical waveform recovery method that can be easily implemented using modern digital-signal-processing technology, achieving high quality real-time waveform recovery and significant reduction of the overall delay, directly from the distorted waveform output from any first-order system with single-pole low-pass transfer characteristics.

## Background

Without loss of generality, I am going to demonstrate the method using an electrometer-grade current-to-voltage amplifier modelled in Fig. 1c which is typically used for scanning tunneling microscopy, non-contact atomic force microscopy, photocurrent detection, etc. Usually its single-pole low-pass characteristics comes from the stray shunt capacitance $C_F$ of the large feedback resistor $R_F$ and we will focus on this case but the actual source of the single-pole characteristics is immaterial. The overall transfer function is then given by:

$$V_O(\omega) = -\frac{R_F||(j\omega C_F)^{-1}}{R_I}V_I(\omega) = -\frac{R_F}{R_I}\frac{1}{1+j\omega R_F C_F}V_I(\omega) \quad (1)$$

In Laplace's s-parameter representation, the transfer function has a single pole at $s = -\frac{1}{R_F C_F}$ ($\tau = R_F C_F$) and no zeros:

$$V_O(s) = -\frac{R_F}{R_I}\frac{1}{1+sR_F C_F}V_I(s) \quad (2)$$

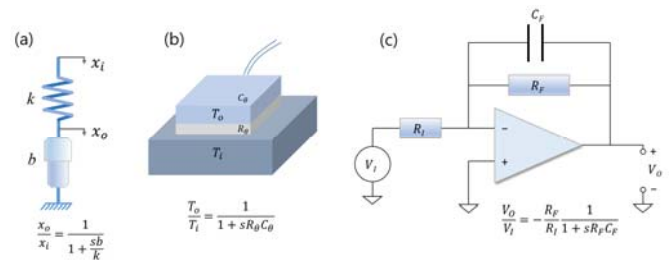

Fig. 1. Examples of sensors and amplifiers with single-pole transfer characteristics. (a) A mechanical system modeled with of a spring ($k$) and a damper ($b$). (b) A thermometer with heat capacity $C_\theta$ and thermal resistance $R_\theta$ between the sample and the thermometer. (c) An electrometer-grade current amplifier with a large feedback resistor $R_F$ and a stray shunt capacitance $C_F$.

[a)]Author to whom correspondence should be addressed. Electronic mail: jhinhwan@kaist.ac.kr

By applying inverse Laplace transform, we can find the corresponding time domain relationship between the input and the output. The s-domain product on the right hand side of Eq. (2) then becomes a time-domain convolution [1]:

$$V_O(t) = -\frac{R_F}{R_I}\int_{-\infty}^{\infty} dt' \left[\frac{e^{-\frac{t'}{\tau}}}{\tau}\Theta(t')\right]V_I(t-t')$$

$$= -\frac{R_F}{R_I}\int_0^{\infty} dt' \frac{e^{-\frac{t'}{\tau}}}{\tau} V_I(t-t') \quad (3)$$

where the normalized convolution kernel $K(t) \equiv \frac{e^{-t/\tau}}{\tau}\Theta(t)$ in the rectangular bracket is an exponential function with time constant $\tau$, multiplied by a step function $\Theta(t)$ required for causality. The value of $\tau$ can be easily determined from a simple step-function-input response measurement for a given sensor or amplifier configuration.

Our goal here is to recover $V_I(t)$ from the measured $\{V_O(t'), t' \leq t\}$ ($t' \leq t$ is required for real-time signal recovery) and a prior knowledge of a particular single-pole convolution kernel $K(t)$. (Note that for a general convolution kernel, the Fourier deconvolution [2,3,4] or several variants of the Richardson-Lucy method [5] can be used but they require a complete input waveform before starting the deconvolution and are thus not suitable for real-time signal recovery. This method can be regarded as a special case where the inverse filter of Ref [6] can be realized.)

**Development for Single-pole case**

Let's compare the following two functions with mutual time domain offset:

$$V_O(t) = -\frac{R_F}{R_I}\int_0^{\infty} dt' \frac{e^{-\frac{t'}{\tau}}}{\tau} V_I(t-t') \quad (4)$$

$$V_O(t-T) = -\frac{R_F}{R_I}\int_0^{\infty} dt' \frac{e^{-\frac{t'}{\tau}}}{\tau} V_I(t-T-t')$$

$$= -\frac{R_F}{R_I}e^{\frac{T}{\tau}}\int_T^{\infty} dt'' \frac{e^{-\frac{t''}{\tau}}}{\tau} V_I(t-t'') \quad (5)$$

Then it is clear that subtracting the first function (Eq. (4)) with the second function (Eq. (5)) multiplied by a factor $\gamma \equiv e^{-\frac{T}{\tau}}$ ($T \ll \tau$) is a good approximation to the original waveform as shown below:

$$V_O(t) - e^{-\frac{T}{\tau}}V_O(t-T) = -\frac{R_F}{R_I}\int_0^T dt' \frac{e^{-\frac{t'}{\tau}}}{\tau} V_I(t-t')$$

$$\approx -\frac{R_F}{R_I}\left(1-e^{-\frac{T}{\tau}}\right)V_I\left(t-\frac{T}{2}\right) \quad (6)$$

$$V_I\left(t-\frac{T}{2}\right) \approx -\frac{R_I}{R_F}\frac{V_O(t)-e^{-\frac{T}{\tau}}V_O(t-T)}{1-e^{-\frac{T}{\tau}}} \equiv -\frac{R_I}{R_F}V_R(t) \quad (7)$$

where

$$V_R(t) = \frac{V_O(t)-\gamma V_O(t-T)}{1-\gamma}. \quad (8)$$

The approximation in Eq. (6) assumes that $V_I(t-t')$ in the integrand can be represented by a $t'$-independent $V_I\left(t-\frac{T}{2}\right)$ over $0 < t' < T$ assuming $V_I$ is a slowly varying function in the time scale of $T$. This is not too restrictive since $T$ can be arbitrarily small and only be limited by the noise performance to be discussed later.

The delayed partial subtraction for real-time evaluation of the recovered signal $V_R(t)$ using Eq. (8) can be performed numerically in a DSP(FPGA) processor as shown in Fig. 2. Here $T = mT_s (\ll \tau)$ is an integer multiple of the sampling period $T_s$ of the DSP(FPGA).

Generally the output $V_R(t)$ of the DSP(FPGA) can reproduce $V_I(t-T_D)$ with maximum delay of $T_D = T + T_{PD}$ where $T_{PD}$ is the total signal propagation delay of the DSP and the AD/DA converters that can be made below 0.1 $\mu$s with a proper selection of modern high-speed and low-latency devices.

Numerical simulations of this waveform recovery process were performed with two different input waveforms of a rectangular wave and a bipolar pulse train as shown in Fig. 3. It is clear that when the delayed correction factor $\gamma$ is exactly equal to $e^{-\frac{T}{\tau}}$, the input signal is perfectly recovered in the final output. While $\tau$ is given by the physical characteristics of the first-order-response analog system, the choice of $T$ can be optimized with the following noise consideration.

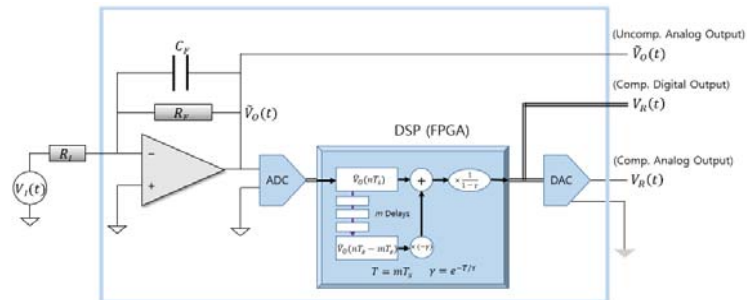

Fig. 2. A stand-alone single-pole-compensated current amplifier. It is effectively an analog amplifier whose frequency characteristics is limited by the feedback resistance's internal $RC$ time constant and the single pole effect is deconvoluted in real time using the moving average process in the DSP(FPGA) equipped with an ADC (and a DAC in case of true stand-alone configuration).

**Noise consideration for Single-pole case**

Due to the numerical "difference" operation $V_O(t) - \gamma V_O(t-T)$ (with $\gamma \approx 1$) involved in the evaluation of $V_R(t)$ in Eq. (8), the simulated noise $V_n(t)$ added to the intermediate output $\tilde{V}_O(t)$ is amplified in the final output $V_R(t)$ as shown in Figs. 3a-b and 3e-f. In order to understand this quantitatively, let's assume without too much loss of generality, that the analog amplifier output signal $\tilde{V}_O(t)$ contains a slow-varying (with respect to $T$) raw signal $V_O(t)$ plus a pseudo-random noise $V_n(t)$ whose correlation time is shorter than $T$. Then the numerical recovery operation applied to $\tilde{V}_O(t) = V_O(t) + V_n(t)$ can be divided into two terms

$$V_R(t) = \frac{V_O(t)-\gamma V_O(t-T)}{1-\gamma} + \frac{V_n(t)-\gamma V_n(t-T)}{1-\gamma} \quad (9)$$

where the first term gives the slow varying signal with value approximated by $V_O(t)$ and the second term gives noise level proportional to $\frac{\sqrt{1+\gamma^2}}{1-\gamma}|V_n(t)|$ due to the presumed absence of time correlation between noise $V_n(t)$ and $V_n(t-T)$. For small $\frac{T}{\tau} \ll 1$ and $\gamma \approx 1$, the signal-to-noise (S/N) ratio is reduced by a factor of

$$\frac{SN_\gamma}{SN_0} \approx \frac{1-\gamma}{\sqrt{1+\gamma^2}} \approx \frac{T}{\sqrt{2}\tau}. \quad (10)$$

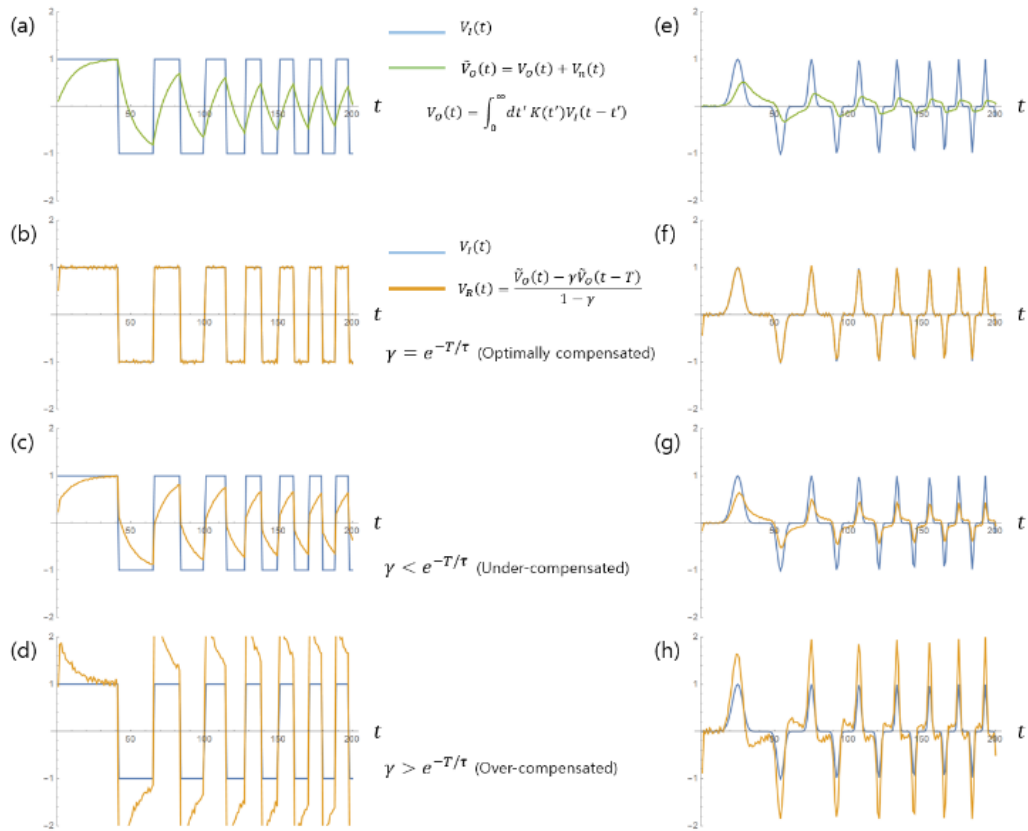

Fig. 3. (Color online) Numerical simulation of the real-time waveform recovery and delay reduction by compensating for a single-pole transfer characteristics. Two example input waveforms of rectangular waves (a-d) and bipolar pulse train (e-h) with decreasing pulse width is used with noise effect simulated with $N_s = 16$. (a) & (e) An intermediate output signal $\tilde{V}_O(t)$ (green) is produced by adding pseudo-random noise $V_n(t)$ to the input signal $V_I(t)$ (blue) convoluted with the single-pole normalized convolution kernel $\frac{e^{-\frac{t}{\tau}}}{\tau}\Theta(t)$. (b) & (f) When the delayed correction factor $\gamma$ is equal to $e^{-\frac{T}{\tau}}$, the recovered output signal $V_R(t) = \frac{\tilde{V}_O(t)-\gamma\tilde{V}_O(t-T)}{1-\gamma}$ matches perfectly well with the original input signal, with increased noise due to the inverse process of the single-pole low-pass filtering. Averaging methods for increasing S/N are suggested in the text. When $\gamma$ is smaller than $e^{-\frac{T}{\tau}}$ ((c) & (g), under-compensated case) or larger than $e^{-\frac{T}{\tau}}$ ((d) & (h), over-compensated case), the signal recovery becomes incomplete and a significant distortion is visible in the final output signal.

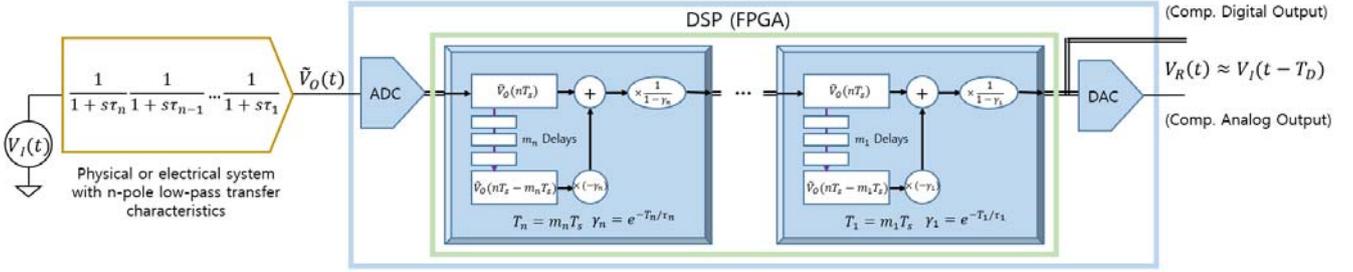

Fig. 4. Diagram of a real-time digital signal recovery system compensating for the signal distortion by a physical or electrical system with *n*-pole low-pass transfer characteristics. The order of removal of the poles is not important.

Therefore, an optimal choice for $T$ can be made such that $T$ should be sufficiently smaller than $\tau$ for the approximation of Eq. (6) to hold but should be increased as much as possible to achieve the largest possible S/N. Fig. 4 shows the cases for $\frac{T}{\tau} = 0.1$, for example. Depending on the source noise level, $0.03 \leq \frac{T}{\tau} \leq 0.3$ would give reasonably good results for a real-time signal recovery required for feedback control system or for a case where repetition of experiment for statistical averaging is not possible.

On the other hand, when an extremely accurate waveform recovery is required even at the expense of a large number $N$ of repeated averaging experiments, a much smaller value of $\frac{T}{\tau} \ll 0.03$ may be used for ultimate accuracy of recovered waveform and yet the final S/N ($SN_\gamma$) can be increased to an arbitrary level by increasing $N$ beyond $N_{min} = 2\frac{\tau^2}{T^2}\left(\frac{SN_\gamma}{SN_0}\right)^2$ since $\frac{SN_\gamma}{SN_0} \approx \frac{T}{\tau}\sqrt{\frac{N}{2}}$.

Even for the real-time signal recovery, if we sample $\tilde{V}_O(t)$ $N_s (\geq 2)$ times over the short time intervals near $t$ and $t - T$ and use their averages in place of $\tilde{V}_O(t)$ and $\tilde{V}_O(t-T)$, we may further increase the S/N by up to a factor given by a fraction of $\sqrt{N_s}$. The factor can approach $\sqrt{N_s}$ in case the correlation time of the noise is still shorter than the sampling periods of $N_s$ data points.

**Development and Noise Considerations for Multi-pole case**

We have shown that the operation $V_R(t) = \frac{V_O(t) - \gamma V_O(t-T)}{1-\gamma}$ removes one s-parameter factor $\frac{1}{1+s\tau}$ from $V_O(s)$ so that $V_R(s) \approx (1+s\tau)V_O(s) = V_I(s)$. Now this operation can be applied $n$ times to recover the original signal from the distorted output of a multipole system with an $n$-pole low-pass transfer function given in the form

$$V_{O_n}(s) = \frac{1}{1+s\tau_n} \cdots \frac{1}{1+s\tau_2} \frac{1}{1+s\tau_1} V_I(s). \quad (11)$$

By assigning intermediate outputs $V_{O_m}(s)$ ($m = 0,1,2,\ldots,n$) with

$$V_{O_0}(s) = V_I(s) \quad (12)$$
$$V_{O_m}(s) = \frac{1}{1+s\tau_m}V_{O_{m-1}}(s) \;(m=1,2,3,\ldots,n), \quad (13)$$

it is clear that the sequential operations

$$\frac{V_{O_m}(t) - \gamma_m V_{O_m}(t-T_m)}{1-\gamma_m} \approx V_{O_{m-1}}(t) \quad (14)$$

with $\gamma_m = e^{-\frac{T_m}{\tau_m}}$ ($m = n, n-1, \ldots, 2, 1$)

will effectively remove all the poles in Eq. (11) one after another and the detailed order of removal is not important. The corresponding implementation diagram is shown in Fig. 4. The total delay will then be the total propagation delay plus the sum of all $T_m$'s or $T_D = T_{PD} + \sum_{m=1}^{n} T_m$. The total S/N will be modified by a factor $\prod_{m=1}^{n} \frac{1-\gamma_m}{\sqrt{1+\gamma_m^2}}\sqrt{N_s} \approx \prod_{m=1}^{n} \frac{T_m}{\tau_m}\sqrt{\frac{N_s}{2}}$ for oversampling by $N_s$.

**Conclusion**

A relatively simple digital-signal-processing-based method of real-time signal recovery is proposed, which can compensate for the waveform distortion and propagation delay due to single-and multi-pole low-pass transfer characteristics in many mechanical, electronic and thermal systems. It will be especially useful in improving the performances of data acquisition systems and stabilizing high speed feedback control systems with sensors and amplifiers operated close to their frequency response limits by utilizing modern low-cost high-speed DSPs and FPGAs.

**Acknowledgements**

This work was supported by the Metrology Research Center Program funded by Korea Research Institute of Standards and Science (No. 2015-15011069), the Pioneer Research Center Program (No. NRF-2013M3C1A3064455), the Basic Science Research Program (No. NRF-2017R1D1A1B01016186) and the Brain Korea 21 Plus Program through the NRF of Korea, and the Samsung Advanced Institute of Technology (SAIT).

**References**

[1]Hazewinkel, Michiel, ed., "Laplace transform" in Encyclopedia of Mathematics, Springer, 2001

[2]Alan V. Oppenheim and Ronald W. Schafer, "Ceptrum Analysis and Homomorphic Deconvolution" in *Discrete-Time Signal Processing*, Englewood Cliffs, NJ, USA: Prentice Hall, 1989

[3]Bracewell, R. N., The Fourier Transform and Its Applications, 3rd ed., Boston: McGraw-Hill, 2000

[4]Steven W. Smith, "Custom Filters" in *The Scientist and Engineer's Guide to Digital Signal Processing*, 1st ed. San Diego, CA, USA: California Technical Publishing, 1997

[5]L. B. Lucy, "An iterative technique for the rectification of observed distributions," Astronomical Journal, vol. 79, pp. 745–754, 1974

[6]D. S.G. Pollock, Richard C. Green and Truong Nguyen, "Linear Filters" in *Handbook of Time Series Analysis, Signal Processing, and Dynamics*, Elsevier, 1999